\documentclass[prb,preprint,showpacs,preprintnumbers,amsmath,amssymb,superscriptaddress]{revtex4}
\usepackage{graphicx}
\usepackage{dcolumn}
\usepackage{bm}

\begin{document}

\title{A biologically-motivated system is poised at a critical
state}

\author{Feng Hu}
\affiliation{College of Physics and Electronic Engineering,
Chongqing Normal University, College Town, Shapingba District,
401331 Chongqing, China.}

\begin{abstract}

We explore the critical behaviors in the dynamics of information
transfer of a biologically-inspired system by an individual-based
model. ``Quorum response'', a type of social interaction which has
been recognized taxonomically in animal groups, is applied as the
sole interaction rule among particles. We assume a truncated
Gaussian distribution to quantitatively depict the distribution of
the particles' vigilance level and find that by fine-tuning the
parameters of the mean and the standard deviation of the Gaussian
distribution, the system is poised at a critical state in the
dynamics of information transfer. We present the phase diagrams to
exhibit that the phase line divides the parameter space into a
super-critical and a sub-critical zone, in which the dynamics of
information transfer varies largely.

\end{abstract}


\date{\today}
\maketitle

% \linenumbers


\section{Introduction}
%%\label{}

Complex system, which is composed of large numbers of components
that interact locally, shares some universal features in a dynamic
process. Self-organized critically (SOC), which was proposed by Bak,
Tang and Wiesenfeld in 1987\cite{SOC paper}, is now a commonly
accepted underlying mechanism to phenomena as earthquakes, solar
quakes, and even dynamics in brain et al\cite{hownatureworks,
earthquake,brainandsoc}. It states that a complex system can
organize itself to a critical state without tuning parameters from
outside. The ``finger print'' of a system entering a critical state
is a power law distribution of the size of the avalanches which is
measured by counting the number of the affected individual
components in the dynamic process. Recently, Cavagana et al.
\cite{Cavagana} observed that in the airborne motion of large
starling flocks, the length of correlation between two individuals'
state doesn't depend on the size of the flock, the so called
scale-free correlation. This observation reveals that the starling
flocks work at a critical state, in which one individual can
effectively affect the state of any others' no matter what the group
size is, and vice versa. This property confers the group an ability
to share information efficiently so that it can optimally respond to
external perturbations. A pioneering study on how information
transfer in a collective animal group was carried out in a fish
school reacting to a risky perturbation in front\cite{Radakov}. It
was found that the fishes at the front made a quick rotation from
the risk and their local neighbors behind imitated this behavior.
The consecutive rotations of the fishes resulted in a rapidly
traveling ``information waves'', which rippled from the front to the
rear at a speed much faster than individual fish's speed. However,
besides these experimental studies, the underlying micro-mechanism
of information transfer is left largely
ignored\cite{Couzin2005,cavaganaThe}. There are some other types of
collective systems, such as swarms of cancer cells\cite{cancercell},
bacterial colonies\cite{Bacterialcolonies} and even human
brains\cite{brain}, share many similarities to the collective animal
groups, and the working efficiency of which may depend on the
underlying mechanism of information transfer.

In this paper, we study the dynamics of information transfer at
critical points in a minimum individual-based model with the
interaction among particles being quorum response. Each particle is
assigned a ``vigilance number'' to quantitatively depict its
vigilance level to respond to its local neighbors' commitment. We
assume the distribution of the ``vigilance number'' to be a
truncated Gaussian distribution in the interval of $(0,1)$. By
tapping information from boundaries into the system, we find that,
by fine tuning parameters of the Gaussian distribution, the system
can be poised at a critical state in the dynamics of information
transfer. We present phase diagrams to show that the critical points
divide the parameter space into a sub-critical and a super-critical
zone, in which the dynamics of information transfer is quite
different.

\section{Quorum response}

Quorum response is a type of social interaction widely found during
the process of collective decision-making in the bee and ant
colonies\cite{bees,ant}, the cockroach aggregations\cite{cockroach},
the broiler chicken crowds\cite{chichens} and the fish
schools\cite{fishes}. It quantitatively states that an individual's
chance of making one option depends on the number of its local
neighbors that committed to this option.

\begin{figure}[ht]
\includegraphics[width=8cm]{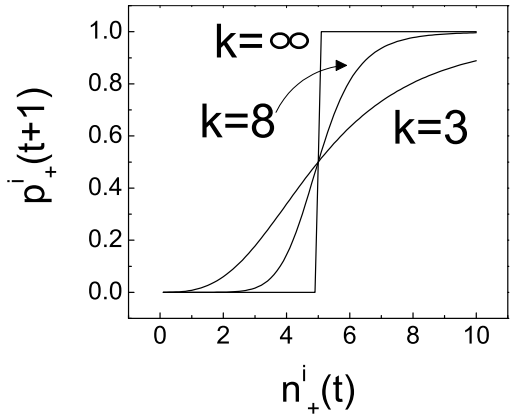}
\vspace{-0.2cm} \caption{Function of quorum response according to
equation (1). The $y$ axis, $p^{i}_{+}(t+1)$, is the probability for
the particle $i$ at time step $t+1$ to choose the option ``+'' and
the $x$ axis, $n^{i}_{+}(t)$, is the number of its local neighbors
that committed to the option ``+'' at time step $t$. The total
number of local neighbors is set to be 10 and the quorum value
$q_{i}$ is set to be 5 in the figure. The parameter of $k=3, 8,
\infty$, which determines the steepness of the curve, is randomly
selected from the infinite series.}
\end{figure}

Let's consider the following simple example, suppose there are only
two options, e.g. being at an alarmed state (``+'' state) or a naive
one (``-'' state). The mathematical description of the rule of
``quorum response'' is as follows\cite{Quorumgeneral}:
\begin{equation}
p^{i}_{+}(t+1)=\frac{(\frac{n^{i}_{+}(t)}{q_{i}})^{k}}{1+(\frac{n^{i}_{+}(t)}{
q_{i}})^{k}},\hspace{0.3cm} p^{i}_{-}(t+1)=1-p^{i}_{+}(t+1)
\end{equation}
Where $p^{i}_{\pm}(t+1)$ is the probability of the individual $i$
choosing to be at an alarmed or a naive state at time step $t+1$,
respectively and $n^{i}_{\pm}(t)$ is the number of local neighbors
who have committed to the alarmed or the naive state at time $t$,
respectively. The term $q_{i}$ is the quorum value for the
individual $i$, which is set to be 5 in Figure 1. We see that the
probability is a monotonic increasing function. Near the quorum
value, it has an inflection point with a rapid increase and the
function is sigmoid. If $k$ is bigger, the variation of the curve
becomes steeper than the linear increase at the quorum value. Thus
it can be expected that $k \geq 2$ is a necessity in the framework
of the interaction rule\cite{Quorumgeneral}. In field experiments,
it is found that the animal group adapt $k$ around
3\cite{cockroach,fishes,Quorumgeneral}. When the parameter $k$
approaches infinity, the plot is practically a step-like switch at
the quorum value, jumping from zero to unity. Quorum response is
essentially a distributed positive feedback process that enables
information propagation and it is believed that this type of
interaction can enhance decision speed and accuracy for a group to
make a collective decision\cite{Quorumgeneral}.

\section{An individual-based Model}

A 2D system is composed of a square of the dimension of
$100\times100$ evenly spaced grid. Each particle is positioned in a
grid, and the individual particle's mobility is ignored because its
speed is far slower than the speed of information propagation
\cite{Radakov}. We assign each particle a ``vigilance number''
$\alpha_{i} \hspace{0.3 cm} (i=1,2,...)$ which measures how vigilant
the particle $i$ is responding to its local neighbor's commitment in
the framework of quorum response. The distribution of $\alpha_{i}$
is assumed to obey a truncated Gaussian distribution in the interval
(0, 1), with the mean being $\mu$ and the standard deviation being
$\Delta$.

The sole interaction rule among the particles in the model is the
quorum response according to equation (1), with the quorum value of
the particle $i$ is defined as,
\begin{eqnarray*}
 q_{i}\equiv n_{0}*\alpha_{i}
\end{eqnarray*}
where $n_{0}=4$ is the constant number of the nearest neighbors to
any particle not positioned at boundaries (if a particle lies at one
of the four corners, $n_{0}=2$, or else if it lies at one of the
four boundary lines, $n_{0}=3$). Each particle can either be in an
alarmed state (``+''state) or in a naive state (``-''state) at the
probability calculated according to equation (1). The probability is
realized by Monte Carlo method at each time step in the dynamic
process of the system, i.e. a random number which is evenly
distributed in the interval of (0,1) is sampled at the time step $t$
and being compared to the probability $p^{i}_{+}(t+1)$ calculated
according to equation (1). If the sampled random number is smaller,
then the particle $i$ will turn into the alarmed state at the next
time step. Otherwise, it will stay in the naive state.

Following the general assumption that individuals at peripheries
find the approaching risks in advance\cite{edgeeffect1}, we tap
information in the system by randomly picking a particle lying at
one of the boundary lines and turn its state to the alarmed one. The
local interactions may affect the state of its neighbor(s), and the
affected neighbors continue to repeat the interaction which may
cascade into an ``information wave'' eventually. If a particle turns
into the alarmed state at time step $t$, then it will stay at the
alarmed state unchanged during the dynamics of information
propagation. One run of information propagation is considered
completed when all the alarmed particles are not capable to alter
the state of its nearest neighbors anymore.

\begin{figure}[ht]
\includegraphics[width=8cm]{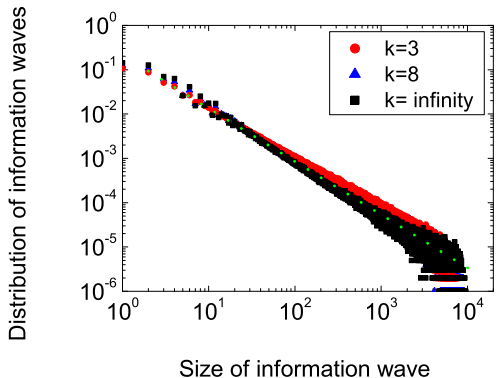}
\vspace{-0.2cm} \caption{Power law distribution of the ``information
waves''. The total population of the particles is $100\times100$ and
the data is averaged over $10^{6}$ runs of simulations. Standard
deviation of the Gaussian distribution is set $\Delta=0.15$ for the
plots. Different $k$ values, paired with particular $\mu$ values are
applied: $k=3$ and $\mu=0.307$ for the red dots, $k=8$ and
$\mu=0.310$ for the blue triangles, $k=\infty$, $\mu=0.316$ for the
black squares. The green dotted straight line, with a slope being
-1.20, is a guide for eyes.}
\end{figure}

We find that if the parameters of $\mu$ and $\Delta$ are fine tuned,
a power law distribution of the size of the information waves is
emerged. The total population of the particles in the system is
$100\times100$ and the standard deviation of the Gaussian
distribution is set to $\Delta=0.15$ in Fig. 2. Each data point is
averaged over $10^{6}$ runs of simulation. The fine tuned parameters
of $\mu$ and $k$ are: $k=3, \mu=0.307; k=8, \mu=0.310$; and
$k=\infty, \mu=0.316$. The size of the information waves is
quantified by counting the number of the particles turned into the
alarmed state. The data collapse on one straight line in a double
logarithmic scale and it is linear for more than three decades with
a slope of -1.20, indicating that the fine-tuned parameters have
poised the system to a critical state\cite{hownatureworks}.

It is interesting to compare the critical behaviors in our model
with its counterpart in the theory of self-organized criticality.
The power law distribution of the size of the information waves, the
counterpart of the dynamical avalanches in SOC, is a similar
``finger print'' to indicate that the system is at a critical state.
The difference is that SOC states that a complex system can organize
itself to a critical state without any tuning of parameters. Yet the
critical state of our model is reached by fine-tuning the
parameters. It was observed recently that not only the bird flocks,
but also some other types of biological systems, e.g. networks of
neurons, even brains, operate near or at the critical points in
parameter space \cite{biocritical,brainandsoc}. It is tempting to
speculate that the fine tuned parameters is a result from a long
adaptation process in the risky nature. When operating at the
critical points, the system becomes more efficient in information
transfer, which confers the system a stronger responsiveness to
predatory attacks.

\section{Phase diagrams}

We explore the relationships of the parameters of $\mu$, $k$ and
$\Delta$ at the critical points, with a population of particles
being $100\times 100$ and the data being obtained over $5\times
10^{5}$ runs of simulations in Fig. 3.

\begin{figure}[ht]
\includegraphics[width=14cm]{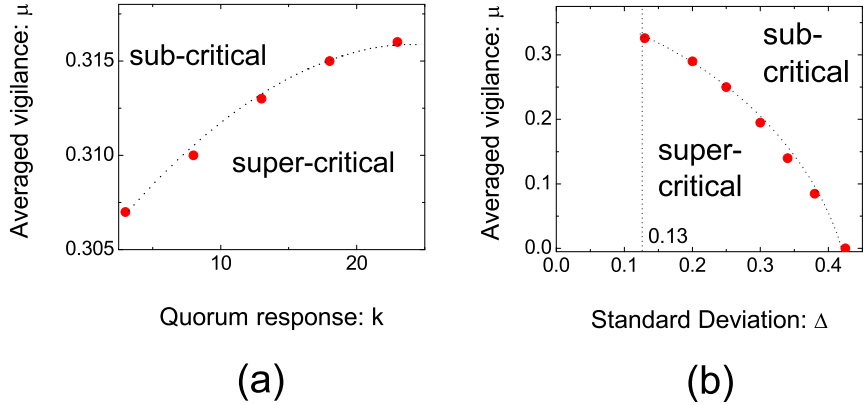}
\vspace{-0.2cm} \caption{Phase diagrams. The population of particles
is $100\times100$ and the data is averaged over $5\times 10^{5}$
runs of simulations The dotted line is a guide for eyes. (a) The $x$
axis is $k$ and $y$ axis is $\mu$, with $\Delta=0.15$ being applied.
The phase line extends to the infinity of $k$ and divides the
parameter space into a super-critical and a sub-critical zone. (b)
The $x$ axis is $\Delta$ and $y$ axis is $\mu$ with $k$ being set to
infinity. The phase line starts from a minimum value of
$\Delta=0.13$ and ends at when $\mu$ is zero. If $\Delta<0.13$, the
system cannot assume a critical state any more. The phase line
divides the parameter space into a super-critical and a sub-critical
zone too.}
\end{figure}

Figure 3(a) shows the phase diagram in the parameter space of $\mu$
and $k$, with $\Delta=0.15$ being applied. For every $k$ extending
from the minimum (restricted by the quorum rules) to infinity, there
is a paired $\mu$ to poise the system to a critical state. As $k$
becomes larger, $\mu$ increases fast initially until it saturates
when $k$ approaches infinity. The phase line divides the parameter
space into a super-critical zone, in which any small perturbations
will very likely cascade into big information waves that affect a
lot of particles, and a sub-critical zone, in which the
perturbations only affect in local areas and information waves are
often blocked by some insensitive particles (with big ``vigilance
number'').

When the system operates at a super-critical state, the random
perturbations of the environment may easily startled the system
because there are so many sensitive particles (with small
``vigilance number''), which will cost a lot of energy waist. On the
contrary, if at a sub-critical state, the big information waves are
damped which results in a low efficiency of information propagation.

Figure 3(b) shows the fine-tuned, paired parameters of $\mu$ and
$\Delta$, which can poise the system at a critical state. Note that
$k$ is set to infinity in the simulations. Along the phase line,
when $\Delta$ becomes bigger, $\mu$ becomes smaller at a rate faster
than the linear drop. There exist a minimum $\Delta=0.13$, where the
phase line starts from. When $\Delta$ is smaller than the minimum,
the system can either be at a sub-critical or a super-critical state
depending on the value of $\mu$, but it cannot reach a critical
state. The phase line ends at $\mu=0$ and $\Delta=0.425$. It is
worth noting that the evenly random distribution of vigilance number
(other than the Gaussian distribution) in the interval $(0, 1)$
results in a sub-critical state.

Since $\Delta$ measures the diversity of the vigilance level of the
group members, Figure 3(b) also shows that if a system is composed
of too many conformists which makes the $\Delta$ value too small,
the system cannot reach a critical state. On the contrary, if the
particle's diversity is too wide, the system cannot reach a critical
state, either.

It has been observed that ants tune the parameters when applying
quorum rules during colony emigration. If the situation is at
different urgent level, e.g. old nest in crisis or in a good
condition, they varied the set of parameters in the situation to
respond to the environment adaptively \cite{q2}.

\section{Conclusion and Discussion}

We proposed a minimum individual-based model with introduction of
quorum response as the local interaction rule to study the critical
behaviors of a biological system, particularly the efficiency of
information transfer. We assumed that the particles' ``vigilance
number'' obeys a truncated Gaussian distribution. By tuning the
parameters of the mean and the standard deviation of the Gaussian
distribution, we found that the system could be poised to a critical
state in a dynamical process of information transfer. We presented
the phase diagrams to show that the parameter space is divided into
a sub-critical and a super-critical zone in which the efficiency of
information propagation vary largely.

In our model, it is assumed that the particle's ``vigilance number''
obeys a same Gaussian distribution. This assumption needs feedback
from experiments, yet the complexity to identify the interaction
among individuals in experiments may stand in the way
currently\cite{Teddy,Katz,Lukeman}. It was observed in field that
individuals of a group of animals at periphery are more vigilant in
average than their conspecifics in center, the so called ``edge
effects''\cite{edgeeffect1}. Although to what degree this effect
works is still under discussion\cite{edgeeffect2}. This effect is
equivalent to positioning particles with relative smaller $q_{i}$ at
the boundaries, which may result in more big information waves thus
enhancing the efficiency of information propagation. The widely
applied interaction rule of collective animal motion in the biology
literatures currently is: ``averaging among the velocity of its
local neighbors''\cite{Vicsek,Buhl,Couzin2005}. Compared with it,
quorum response has the advantages of passing information on without
deterioration. This may be the reason that in some risky
environments animal groups apply quorum response rules to make the
movement decision \cite{fishes} and even tuning the parameters if in
urgency \cite{q2}.

Collective animal groups, e.g. starling flocks, fish schools, are
recently being called ``collective minds''\cite{Collectiveminds},
because the coherent movements and the efficient responses to
environmental perturbations as if the whole group is in one mind.
Obviously, the enhanced efficiency of information transfer in the
group enables these abilities. Our model explores the dynamics of
information transfer in a system at critical points. Hopefully, our
work may shed some light on this mysterious and beautiful
phenomenon.
\bibliographystyle{elsarticle-num}

\end{document}